\documentclass[preprint]{aastex}

\begin{document}

\title{Main-Belt Comet P/2012 T1 (PANSTARRS)
\footnote{
Some of the data presented herein were obtained at the W.\ M.\ Keck Observatory, which is operated as a scientific partnership among the California Institute of Technology, the University of California, and the National Aeronautics and Space Administration, and made possible by the generous financial support of the W.\ M.\ Keck Foundation, the Magellan Telescopes located at Las Campanas Observatory, Chile, and the Southern Astrophysical Research (SOAR) telescope, which is a joint project of the Minist\'{e}rio da Ci\^{e}ncia, Tecnologia, e Inova\c{c}\~{a}o (MCTI) da Rep\'{u}blica Federativa do Brasil, the U.S. National Optical Astronomy Observatory (NOAO), the University of North Carolina at Chapel Hill (UNC), and Michigan State University (MSU).
}
}

\author{
Henry H.\ Hsieh\altaffilmark{1,a},
Heather M.\ Kaluna\altaffilmark{1,b},~
Bojan Novakovi\'c\altaffilmark{2},~
Bin Yang\altaffilmark{1,b},~
Nader Haghighipour\altaffilmark{1,b},~
Marco Micheli\altaffilmark{1},~
Larry Denneau\altaffilmark{1},~
Alan Fitzsimmons\altaffilmark{3},~
Robert Jedicke\altaffilmark{1},~
Jan Kleyna\altaffilmark{1,b},~
Peter Vere{\v s}\altaffilmark{1},~
Richard J.\ Wainscoat\altaffilmark{1},~
Megan Ansdell\altaffilmark{1,b},~
Garrett T.\ Elliott\altaffilmark{1},~
Jacqueline V.\ Keane\altaffilmark{1,b},~
Karen J.\ Meech\altaffilmark{1,b},~
Nicholas A.\ Moskovitz\altaffilmark{4},~
Timm E.\ Riesen\altaffilmark{1,b},~
Scott S.\ Sheppard\altaffilmark{5},~
Sarah Sonnett\altaffilmark{1},~
David J.\ Tholen\altaffilmark{1},~
Laurie Urban\altaffilmark{1,b},~
Nick Kaiser\altaffilmark{1},~
K.~C.\ Chambers\altaffilmark{1},~
William S. Burgett\altaffilmark{1},~
Eugene A.\ Magnier\altaffilmark{1},~
Jeffrey S. Morgan\altaffilmark{1},~
Paul A.\ Price\altaffilmark{6}~
}

\altaffiltext{1}{Institute for Astronomy, University of Hawaii, 2680 Woodlawn Drive, Honolulu, HI 96822, USA}
\altaffiltext{2}{Department of Astronomy, Faculty of Mathematics, University of Belgrade, Studentski trg 16, 11000 Belgrade, Serbia}
\altaffiltext{3}{Astrophysics Research Centre, Queens University Belfast, Belfast BT7 1NN, United Kingdom}
\altaffiltext{4}{Department of Earth, Atmospheric, \& Planetary Sciences, Massachusetts Institute of Technology, 77 Massachusetts Avenue, Cambridge, MA 02139, USA}
\altaffiltext{5}{Department of Terrestrial Magnetism, Carnegie Institution of Washington, 5241 Broad Branch Road, NW, Washington, DC 20015, USA}
\altaffiltext{6}{Department of Astrophysical Sciences, Peyton Hall, Princeton University, Princeton, 08544, USA}
\altaffiltext{a}{Hubble Fellow}
\altaffiltext{b}{NASA Astrobiology Institute}
\email{hsieh@ifa.hawaii.edu}

\slugcomment{Submitted, 2013-04-25; Accepted, 2013-05-22}

\begin{abstract}
We present initial results from observations and numerical analyses aimed at characterizing main-belt comet P/2012 T1 (PANSTARRS).  Optical monitoring observations were made between October 2012 and February 2013 using the University of Hawaii 2.2~m telescope, the Keck I telescope, the Baade and Clay Magellan telescopes, Faulkes Telescope South, the Perkins Telescope at Lowell Observatory, and the Southern Astrophysical Research (SOAR) telescope.  The object's intrinsic brightness approximately doubles from the time of its discovery in early October until mid-November and then decreases by $\sim$60\% between late December and early February, similar to photometric behavior exhibited by several other main-belt comets and unlike that exhibited by disrupted asteroid (596) Scheila.  We also used Keck to conduct spectroscopic searches for CN emission as well as absorption at 0.7~$\mu$m that could indicate the presence of hydrated minerals, finding an upper limit CN production rate of $Q_{\rm CN}<1.5\times10^{23}$~mol~s$^{-1}$, from which we infer a water production rate of $Q_{\rm H_2O}<5\times10^{25}$~mol~s$^{-1}$, and no evidence of the presence of hydrated minerals.  Numerical simulations indicate that P/2012 T1 is largely dynamically stable for $>100$~Myr and is unlikely to be a recently implanted interloper from the outer solar system, while a search for potential asteroid family associations reveal that it is dynamically linked to the $\sim$155~Myr-old Lixiaohua asteroid family.\end{abstract}

\keywords{comets: general ---
          minor planets, asteroids: general}

\newpage

\section{INTRODUCTION}

\begin{figure}
\plotone{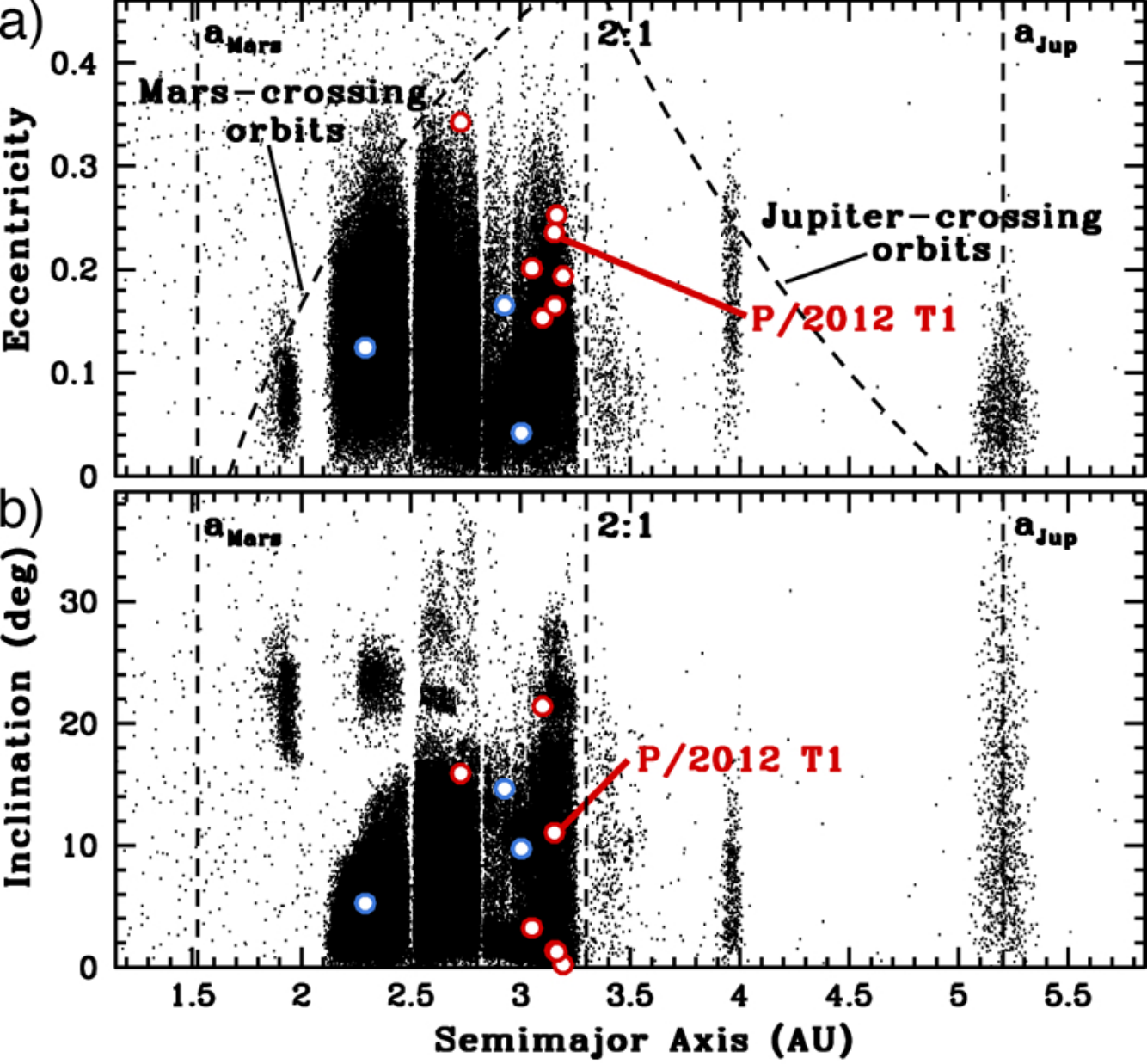}
\caption{\small Semimajor axis versus (a) eccentricity and (b) inclination plots for probable MBCs (red circles) and disrupted asteroids (blue circles).
}
\label{fig_aeimbcs}
\end{figure}

Main-belt comets \citep[MBCs;][]{hsi06} exhibit cometary activity indicative of sublimating ice, yet orbit entirely within the main asteroid belt (Figure~\ref{fig_aeimbcs}).  Seven MBCs --- 133P/Elst-Pizarro, 176P/LINEAR, 238P/Read, 259P/Garradd, P/2010~R2 (La Sagra), P/2006~VW$_{139}$, and P/2012~T1 (PANSTARRS) --- are currently known.  In addition, three other objects --- P/2010~A2 (LINEAR), (596) Scheila, and P/2012~F5 (Gibbs) --- have been observed to exhibit comet-like dust emission, though their active episodes have been attributed to impact events and are not believed to be sublimation-driven \citep{jew10,jew11,sno10,bod11,ish11,ste12}.  As such, we do not consider these objects to be ice-bearing main-belt objects, and refer to them as disrupted asteroids (Figure~\ref{fig_aeimbcs}).

\section{OBSERVATIONS}

P/2012 T1 was discovered on 2012 October 6 by the 1.8~m Pan-STARRS1 (PS1) survey telescope on Haleakala \citep{wai12}.  PS1 employs a $3.2\degr\times3.2\degr$ 1.4 gigapixel camera, consisting of 60 orthogonal transfer arrays, each comprising 64 $590\times598$~pixel CCDs.  Our discovery observations were made using Sloan Digital Sky Survey (SDSS) $r'$- and $i'$-like filters designated $r_{\rm P1}$ and $i_{\rm P1}$ \citep{ton12}.  Comet candidate identification in PS1 data is accomplished using automated point-spread function (PSF) analysis procedures \citep{hsi12b} implemented as part of PS1's Moving Object Processing System \citep[MOPS;][]{den13}.

Follow-up observations were obtained in photometric conditions between October 2012 and February 2013 using
the University of Hawaii (UH) 2.2~m and the 10~m Keck I telescopes, both on Mauna Kea,
the 6.5~m Baade and Clay Magellan telescopes at Las Campanas,
the 2.0~m Faulkes Telescope South (FTS) at Siding Spring,
the 1.8~m Perkins Telescope (PT) at Lowell Observatory,
and the Southern Astrophysical Research (SOAR) telescope on Cerro Pachon
(Table~\ref{table_obslog}; Figure~\ref{fig_observations}a,b).
We employed
a 2048$\times$2048 pixel Textronix CCD for UH observations,
the Low Resolution Imaging Spectrometer \citep[LRIS;][]{oke95} for Keck observations,
the Inamori Magellan Areal Camera and Spectrograph (IMACS) for Baade observations,
the Megacam mosaic camera (consisting of 36 2048$\times$4608 pixel CCDs) for Clay observations,
a 4096$\times$4096 pixel Fairchild CCD for FTS observations,
the Perkins ReImaging System for Lowell observations,
and the SOAR Optical Imager \citep[SOI;][]{sch04} for SOAR observations.
We used SDSS-like filters for Clay observations, Bessell filters for FTS observations, and Kron-Cousins filters for all other observations.
UH 2.2~m, Keck, Lowell, and SOAR observations were conducted using non-sidereal tracking at the apparent rate and direction of motion of P/2012 T1 on the sky, while other observations were conducted using sidereal tracking.

PS1 data were reduced using the system's Image Processing Pipeline \citep[IPP;][]{mag06}.  We performed bias subtraction and flat-field reduction for follow-up data using Image Reduction and Analysis Facility \cite[IRAF;][]{tod86} software and using flat fields constructed either from images of the illuminated interior of the telescope dome or dithered images of the twilight sky.  Some photometric calibration was performed using field star magnitudes provided by the Sloan Digital Sky Survey \citep[SDSS;][]{yor00} converted to Kron-Cousins $R$-band equivalents using the transformation equations derived by R.\ Lupton (available online at {\tt http://www.sdss.org/}).  Photometry of \citet{lan92} standard stars and field stars was performed for all data using IRAF and obtained by measuring net fluxes within circular apertures, with background sampled from surrounding circular annuli.  Conversion of $r'$-band magnitudes measured from PS1 and Clay data to their $R$-band equivalents was performed assuming approximately solar colors for the object.

Comet photometry was performed using circular apertures, where to avoid dust contamination from the comet itself, background sky statistics are measured manually in regions of blank sky near, but not adjacent, to the object.  Photometry aperture sizes were chosen to encompass $>$95\% of the total flux from the comet (coma and tail) while minimizing interference from nearby field stars or galaxies, and varied from $3\farcs0$ to $10\farcs0$ in radius depending on seeing conditions.  Field stars in comet images were also measured to correct for any extinction variations during each night.

In addition to imaging, we also obtained optical spectra of P/2012 T1 on 2012 October 19 with LRIS in spectroscopic mode on Keck.  Two G2V solar analog stars, HD28099 and HD19061, were also observed to allow removal of atmospheric absorption features and calculation of P/2012 T1's relative reflectance spectrum. We utilized a $1\farcs0$-wide long-slit mask and LRIS's D500 dichroic, with a 400/3400 grism on the blue side (dispersion of 1.09~\AA~pixel$^{-1}$ and spectral resolution of $\sim$7~\AA), and 150/7500 grating on the red side (dispersion of 3.0~\AA~pixel$^{-1}$ and spectral resolution of $\sim$18~\AA).  Exposure times totaled 1320~s and 1200~s on the blue and red sides, respectively, where the comet was at an airmass of $\sim$1.2 during our observations. Data reduction was performed using IRAF.

\section{RESULTS \& ANALYSIS\label{results}}

\subsection{Photometric Analysis\label{photresults}}

Photometry results from follow-up observations are listed in Table~\ref{table_obslog}.  For reference, we also compute $Af\rho$ \citep{ahe84} for each of our observations, though we note that it is not always a reliable measurement of the dust contribution to comet photometry in cases of non-spherically symmetric comae \citep[e.g.,][]{fin13}.

While much of our photometry are based on snapshot observations (meaning that unknown brightness variations due to nucleus rotation could be present), we find that the object's intrinsic brightness roughly doubles from the time of its discovery in early October until mid-November ($\sim$40 days; over a true anomaly range of $7^{\circ}<\nu<20^{\circ}$), and then decreases by $\sim$60\% between late December and early February ($\sim$50 days; $28^{\circ}<\nu<42^{\circ}$) (Figure~\ref{fig_observations}c).  Similar photometric behavior is observed for several other MBCs \citep{hsi12b,hsi12c}.  For comparison, the brightness of disrupted asteroid (596) Scheila declined by 30\% in just 8 days \citep{jew11}.  MBCs 133P and 238P both exhibited long-lived brightening and did so during multiple apparitions, making us extremely confident in their cometary natures \citep{hsi04,hsi10,hsi11}.  While long-lived activity is no guarantee of cometary activity \citep{hsi12a}, the photometric behavior of P/2012 T1 is certainly inconsistent with dust particles ejected impulsively in an impact.  Its steady brightening implies the action of a prolonged dust ejection mechanism like sublimation.  Furthermore, while apparently long-lived activity could be due to the long dissipation times of large particles ejected by an impact, P/2012~T1's eventual fading after several weeks suggests that this is not the case here, since such large particles would be expected to persist much longer \citep[cf.][]{hsi04,jew10}.

Multi-filter observations using LRIS on Keck I (which permits simultaneous $B$- and $R$-band imaging, eliminating the effects of rotational brightness variations) and the Baade telescope indicates that coma of P/2012 T1 had approximately solar colors of $B-R=1.13\pm0.04$ (measured on Keck), and $B-V=0.65\pm0.07$ and $V-R=0.37\pm0.05$ (measured on Baade).

\subsection{Spectroscopic Analysis\label{specresults}}

\begin{figure}
\plotone{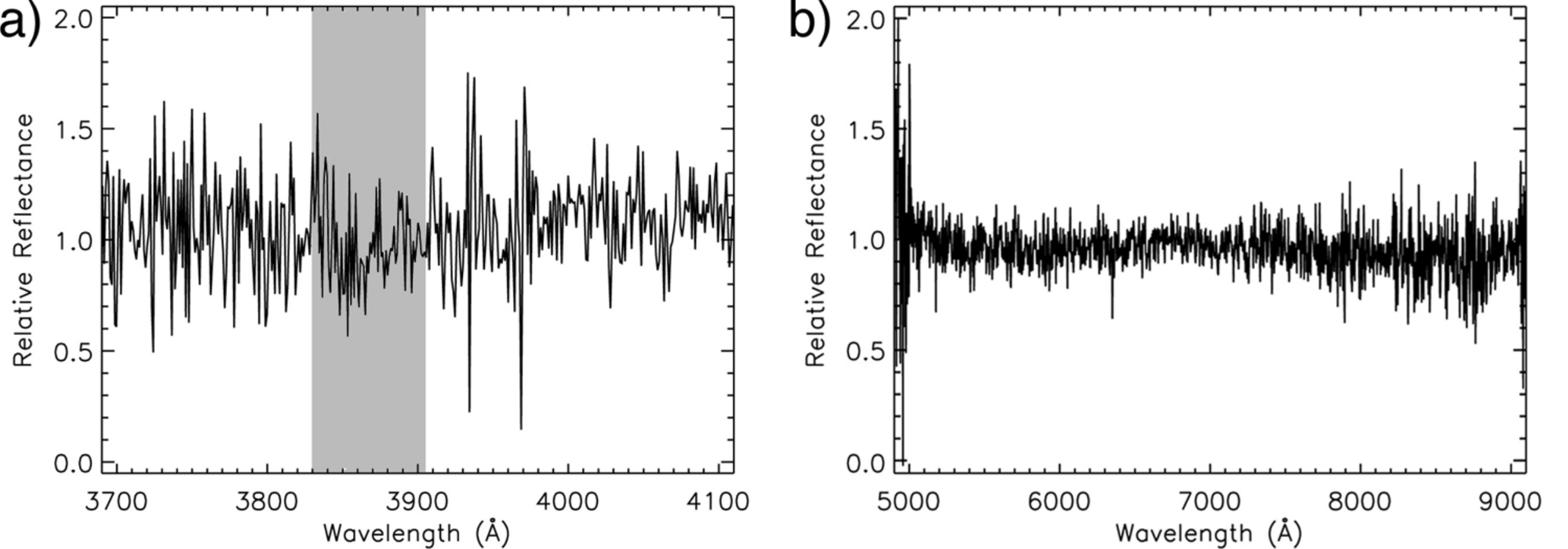}
\caption{\small Relative reflectance spectra of P/2012 T1 obtained with LRIS on Keck I on 2012 October 19 from (a) 3700 \AA\ to 4100 \AA, where the shaded region indicates the wavelength region where the CN emission band is expected, and (b) 5000 \AA\ to 9000 \AA, where absorption at 0.7$\mu$m due to a charge transfer transition in oxidized iron is expected if hydrated minerals are present.
}
\label{fig_spectra}
\end{figure}

Our LRIS red-side spectrum (Figure~\ref{fig_spectra}) of P/2012 T1 is approximately linear with a slightly blue slope of $-1.5\pm1.0$~\%/1000\AA, similar to the spectrum of 133P when it was active during its 2007 perihelion passage \citep{lic11}.  This result differs significantly, however, from the red slopes measured for MBC P/2006 VW$_{139}$ when it was active \citep[7.2\%/1000\AA;][]{hsi12b} as well as for other cometary dust comae \citep{kol04}.

To derive the CN production rate \citep[cf.][]{hsi12a,jew12a}, we employ a simple \citet{has57} model, using a resonance fluorescence efficiency of $g[1{\rm AU}]=3.63\times10^{-13}$~erg~s$^{-1}$~molecule$^{-1}$ \citep{sch10}.  We find an upper limit to the CN production rate of $Q_{\rm CN}<1.5\times10^{23}$~mol~s$^{-1}$.  The CN to water production rate in MBCs is unknown, but we adopt the average ratio in other observed comets ($\log [{Q_{\rm CN}}/{Q_{\rm OH}}]\sim-2.5$; ${Q_{\rm OH}}/{Q_{\rm H_{2}O}}\sim90$\%) \citep{ahe95}, and infer a water production rate of $Q_{\rm H_{2}O}<5\times10^{25}$~mol~s$^{-1}$.

We also search for 0.7~$\mu$m absorption due to a charge transfer transition in oxidized iron in phyllosilicates, indicative of the presence of hydrated minerals \citep{vil94}.  Thermal evolution models suggest that aqueous alteration occurred within asteroid parent body interiors \citep{coh00,wil99}. If these models are correct, MBCs could be icy fragments from the outer shells of asteroid parent bodies where temperatures were never high enough to melt ice.  If an MBC happened to be a fragment from near an ice and hydrated rock boundary in such a parent body, it could contain hydrated minerals.  To date, no MBCs have shown evidence of having hydrated minerals on their surfaces.  Our Keck spectrum of P/2012 T1 likewise shows no signs of absorption at 0.7~$\mu$m, and thus, no detectable evidence of hydrated minerals.

\subsection{Dynamical Analysis\label{dynresults}}

\subsubsection{Stability Analysis\label{stability}}

To determine whether P/2012 T1 is likely to be native to the main belt, or if it could be a recently implanted interloper from the outer solar system, we analyze its long-term dynamical stability in a manner similar to that performed for other MBCs \citep[cf.][]{jew09,hsi12a,hsi12b,hsi12c}.  We generate nine sets of 100 dynamical clones of P/2012 T1 that are Gaussian-distributed in orbital element space and centered on the object's osculating orbital elements as of 2012 December 1.  Three of these sets are characterized by $\sigma$ values equivalent to the uncertainties on those orbital elements ($\sigma_a=3\times10^{-5}$~AU, $\sigma_e=3\times10^{-5}$, $\sigma_i=6^{\circ}\times10^{-4}$), three sets are characterized by $\sigma$ values 10 times larger than those uncertainties, and three sets are characterized by $\sigma$ values 100 times larger.  We then perform forward integrations for 100 Myr using the N-body integration package, Mercury \citep{cha99}.  We include the gravitational effects of all eight major planets and treat all dynamical clones as massless test particles.  Non-gravitational forces are not considered in this analysis.

\begin{figure}
\plotone{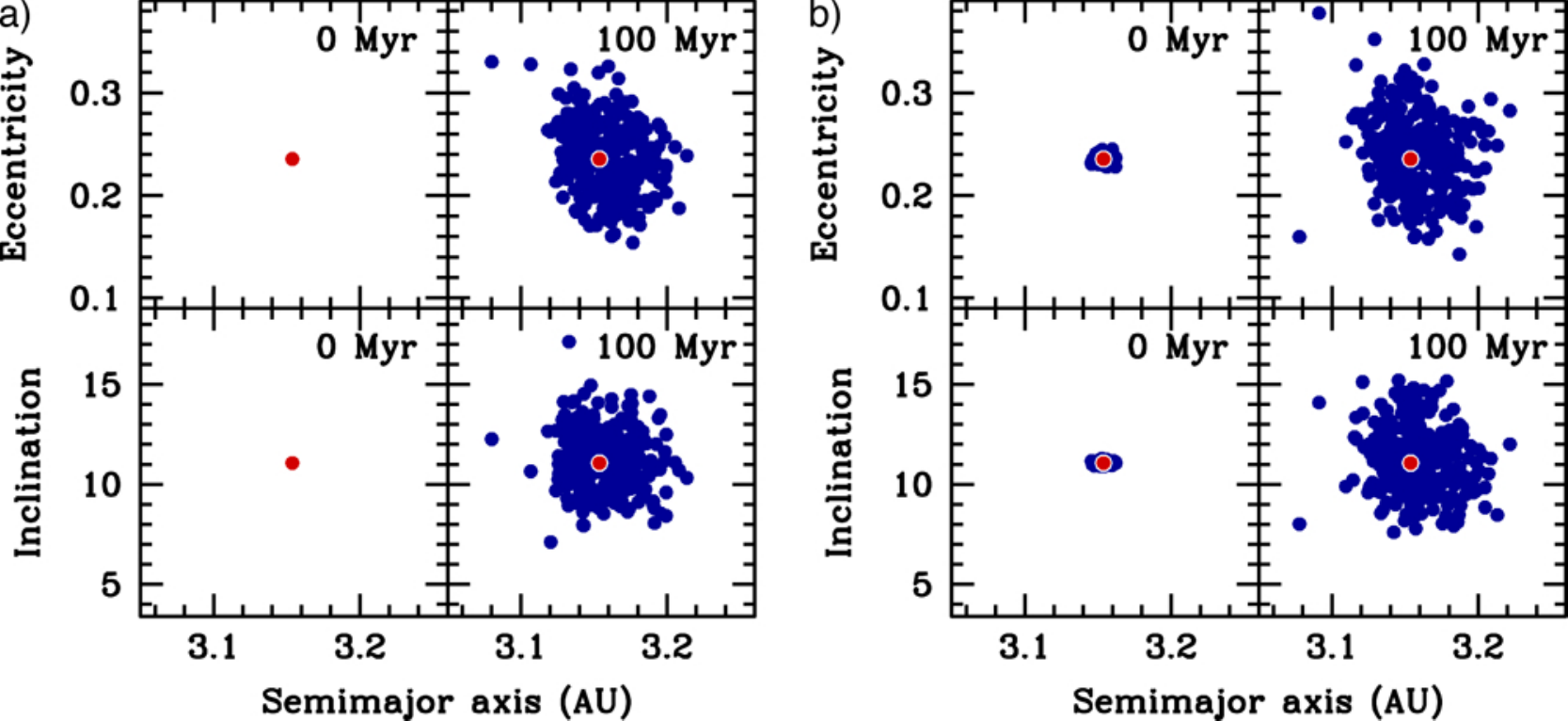}
\caption{\small Semimajor axis vs.\ eccentricity and inclination plots for (a) all 1-$\sigma$ sets of dynamical clones and (b) all 100-$\sigma$ sets of clones of P/2012 T1 integrated as described in Section~\ref{stability}, where the orbital elements of all clones (blue dots) are shown at the beginning (0~Myr) and at the end (100~Myr) of each integration.  For reference, the original orbital elements of the object are marked with a red dot in each plot.
}
\label{fig_stability}
\end{figure}

In these simulations, less than 5\% of the test particles reach heliocentric distances of $>50$~AU (and are therefore considered to have been ejected from the asteroid belt) over the 100~Myr test period.  The remaining test particles diverge to occupy regions of orbital element space that are larger than their initial distributions but that are also essentially independent of those initial distributions, i.e., the 1-$\sigma$ sets of test particles diverge to occupy similar regions as the 100-$\sigma$ sets (Figure~\ref{fig_stability}).  This divergence occurs quickly (within $10^4$~years) and remains approximately constant over the 100~Myr test period.  We therefore find that P/2012 T1 is largely dynamically stable and is unlikely to be a recently implanted interloper, though we do note that the ejection of a small number of test particles indicates that the region is not perfectly stable over the considered time period.

\subsubsection{Search for Associated Dynamical Families\label{family}}

MBCs 133P and P/2006 VW$_{139}$ have recently been found to be dynamical members of very young ($<$10~Myr) asteroid families \citep{nes08,nov12}.  These findings are interesting because the surfaces of ice-bearing main-belt objects may become significantly collisionally devolatilized on timescales of $\ll1$~Gyr \citep{hsi09}.  However, if MBCs only originated in the recent fragmentation events that created the aforementioned young families, their surfaces should have experienced far less collisional devolatilization, and thus remain susceptible to activation by small impactors \citep[cf.][]{hsi04,cap12}.  While we currently lack a sufficient sample size to ascertain whether there is a significant overabundance of MBCs in young families, the fact that two of seven MBCs ($\sim$30\%) are found to belong to such families is suggestive of a physical correlation.

To test whether P/2012 T1 originated in a recent fragmentation event, we search for an associated dynamical family utilizing a hierarchical clustering analysis \citep{zap90}.  Using computed proper elements of $a_p=3.15967$~AU, $e_p=0.19555$, and $\sin(i_p)=0.17536$, we find that P/2012~T1 belongs to the $\sim$155~Myr-old Lixiaohua asteroid family \citep{nov10}.  While the Lixiaohua family is of intermediate age, P/2012~T1 could belong to an even younger sub-family, much as the young Beagle and P/2006 VW$_{139}$ families are both sub-groups of the much older Themis family \citep{nes08,nov12}.  Unfortunately, the density of asteroids in the region of orbital element space occupied by the Lixiaohua family is extremely high.  In fact, most Lixiaohua family members, including P/2012 T1, are linked to the family at cut-off velocities as low as 20~m~s$^{-1}$.  As such, identification of a younger sub-family for P/2012 T1 will be extremely difficult.  The slight instability of the P/2012 T1's orbit (Section~\ref{stability}) also interferes with our ability to establish young family linkages using techniques such as those applied in the case of P/2006 VW$_{139}$ \citep{nov12}.

\section{DISCUSSION\label{discussion}}

Currently, the key question that must be answered when a main-belt object exhibits comet-like activity is whether that activity is sublimation-driven, implying the presence of ice, or is produced by another means.  Definitive evidence of sublimation would be the direct detection of a gaseous sublimation product such as CN or H$_2$O in the coma of such an object.  Unfortunately, unsuccessful attempts to detect CN have now been made for four of the most recently discovered MBCs \citep[][this work]{jew09,hsi12b,hsi12c,lic13}, where each work has found similar upper limit CN production rates of $10^{23}-10^{24}$~mol~s$^{-1}$, corresponding to water production rates of $Q_{\rm H_2O}<10^{26}$~mol~s$^{-1}$.  Searches for line emission from the ($1_{10}-1_{01}$) rotational transition of ortho-water at 557 GHz with the {\it Herschel} Space Observatory for 176P and P/2012 T1 were also unsuccessful, finding $Q_{\rm H_2O}<4\times10^{25}$~mol~s$^{-1}$ and $Q_{\rm H_2O}<7\times10^{25}$~mol~s$^{-1}$, respectively \citep[][O'Rourke et al., private communication]{dev12}.  While these results do not definitively rule out sublimation, they do indicate that the production rates of sublimation products by MBCs are too low to detect from current Earth-bound facilities.  As such, we must rely on indirect evidence to determine the likely source of comet-like activity in main-belt objects.

\citet{jew12} examined various mechanisms by which an asteroid-like body could undergo comet-like mass loss, including ice sublimation, impact ejection, rotational instability, and electrostatic levitation, finding that in many individual cases of comet-like objects, the cause of observed mass loss could not be definitively identified due to insufficient observational data.  Nonetheless, certain mechanisms could sometimes be ruled out based on physical and observational constraints.

For example, electrostatic levitation was ruled out as a cause of 133P's observed activity because it would have depleted the supply of mobile surface dust during a single active episode, leaving no obvious source of mobile dust for subsequent active episodes \citep{hsi04,hsi10}.  The rapid rotation of 133P also minimizes the amount of electrostatic charging that can occur given the short time that any portion of the object's surface spends in sunlight \citep{hsi04}.  This mechanism's efficacy furthermore depends on unknown cohesive properties of asteroid regolith dust grains \citep{jew12}.  Finally, given the many asteroids similar to 133P with more favorable rotational properties, it is unclear why 133P would exhibit observable dust levitation while other asteroids do not.  Dust ejection via rotational spin-up, perhaps via the Yarkovsky-O'Keefe-Radzievsky-Paddack (YORP) effect \citep{rub00}, was also ruled out due to the lack of a plausible mechanism for producing repeated activity or explaining the rarity of similar activity on other asteroids.

\citet{jew12} did note that for an object exhibiting repeated activity, sublimation appears to be the only reasonable explanation.  Repeated activity has only been established for 133P and 238P, however, with others either failing to exhibit repeated activity upon completion of a full orbit period since its previously observed active episode (176P), or not yet having completed one full orbit since their first observed active episodes (259P, P/2010 R2, P/2006 VW$_{139}$, and P/2012 T1).

\begin{figure}
\plotone{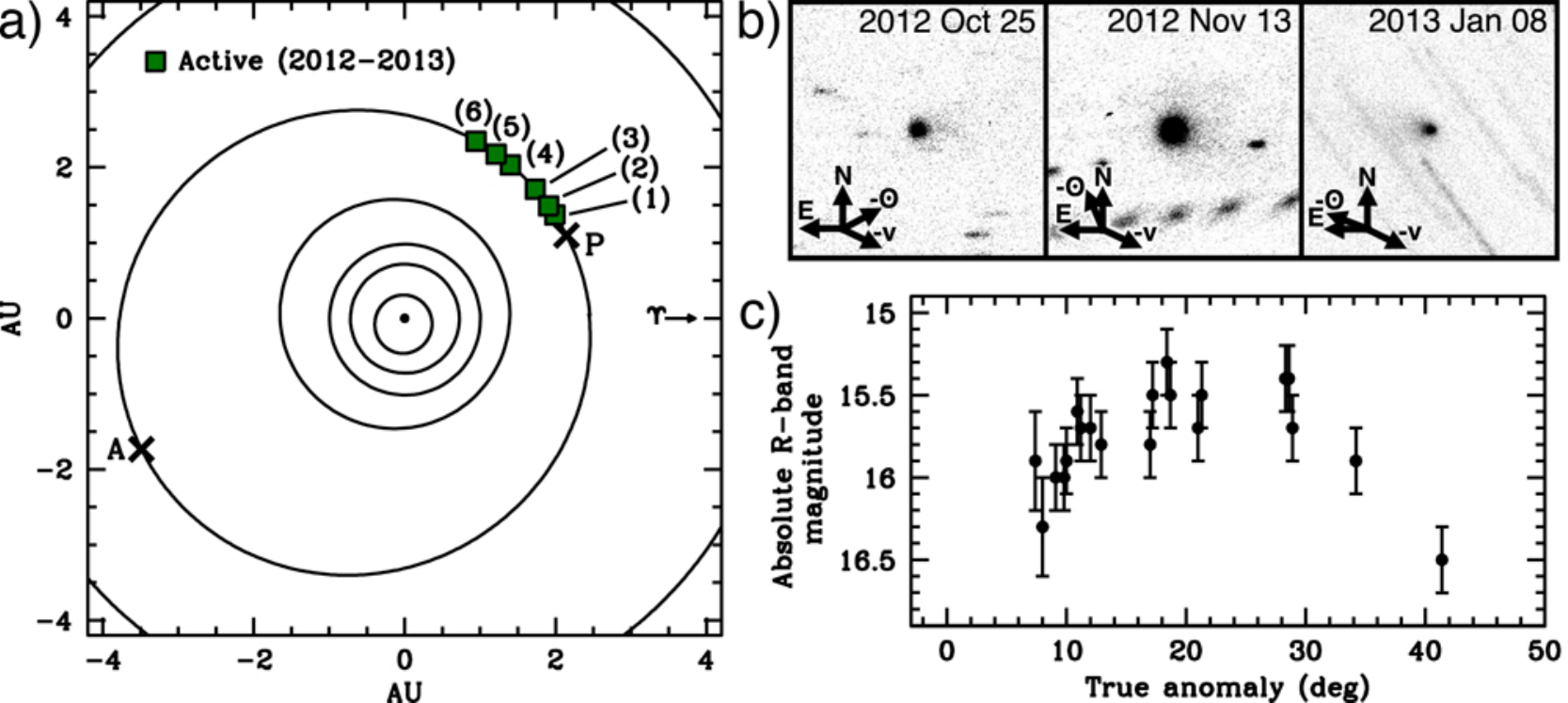}
\caption{\small (a) Orbital position plot with the Sun (black dot) at the center, and the orbits of Mercury, Venus, Earth, Mars, P/2012 T1, and Jupiter shown as black lines. Perihelion (P) and aphelion (A) are marked with crosses.  Green squares correspond to observations from (1) 2012 October 6-8, (2) 2012 October 12-25, (3) 2012 November 8-14, (4) 2012 December 18-20, (5) 2013 January 8, and (6) 2013 February 4.
(b) Composite images of P/2012 T1 (center of each panel).  In each panel, North (N), East (E), and the antisolar ($-\odot$) and negative heliocentric velocity ($-v$) directions are marked.
(c) Plot of absolute magnitude versus true anomaly for observations listed in Table~\ref{table_obslog}.
}
\label{fig_observations}
\end{figure}

As discussed in Section~\ref{photresults}, P/2012~T1's observed photometric behavior indicates ongoing and even increasing dust production over several weeks (Figure~\ref{fig_observations}c), consistent with continuous sublimation-driven dust ejection and inconsistent with impulsive impact-driven dust ejection.  The comet also exhibits a diffuse coma and a featureless fan-like tail that remains aligned with the antisolar direction, distinctly different from the crossed filamentary structure of P/2010 A2's tail, the three-plumed morphology of (596) Scheila's dust tail, and the orbit-plane-aligned dust trail of P/2012 F5 \citep{jew10,ish11,ste12,mor12}.  The post-perihelion peaking of P/2012 T1's activity is also consistent with the post-perihelion peaking of activity for other MBCs \citep{hsi12c}.  While we cannot yet definitively conclude that P/2012 T1's activity is sublimation-driven, we note that all evidence examined thus far is consistent with sublimation.  We therefore find that P/2012 T1 is most likely a true MBC, and not a disrupted asteroid, though additional observations (e.g., to search for repeated activity during its next perihelion passage in mid-2018) and more detailed dust modeling will be required to definitively rule out other dust ejection mechanisms.

A primary ultimate objective of MBC studies is to connect observations of the distribution and composition of volatiles in small primitive bodies to the distribution of volatiles in the protoplanetary disk, and to link this through disk observations to other forming planetary systems \citep{pon10}. Presently, we have insufficient information to make these connections, in part because we have few direct constraints on the volatile content of small bodies and because our solar system's precise dynamical history remains poorly understood.  Further work on both fronts would help this situation, though significantly advancing our understanding of volatile material in the asteroid belt may require in-situ investigation, e.g., by a visiting spacecraft.

\begin{acknowledgements}
H.H.H.\ is supported by NASA through Hubble Fellowship grant HF-51274.01 awarded by the Space Telescope Science Institute, which is operated by the Association of Universities for Research in Astronomy, Inc., for NASA, under contract NAS 5-26555.
H.M.K., B.Y., N.H., and K.J.M.\ acknowledge support through the NASA Astrobiology Institute under Cooperative Agreement NNA09DA77A.
B.N. is supported by the Ministry of Education and Science of Serbia under Project 176011.
J.K., M.A., J.V.K., T.R., and L.U.\ acknowledge support through NSF grant AST-1010059.
We thank Larry Wasserman and Brian Taylor at Lowell for assistance in obtaining observations.
Some data presented was acquired using the PS1 System operated by the PS1 Science Consortium (PS1SC) and its member institutions.  The PS1 survey was made possible by contributions from PS1SC member institutions and NASA through Grant NNX08AR22G issued through the Planetary Science Division of the NASA Science Mission Directorate.
SDSS-III (http://www.sdss3.org/) is funded by the Alfred P.\ Sloan Foundation, the Participating Institutions, NSF, and the U.S.\ Department of Energy Office of Science, and managed by the Astrophysical Research Consortium for the SDSS-III Collaboration.
\end{acknowledgements}

\begin{deluxetable}{lcrrcrrrrrrrccc}
\tabletypesize{\tiny}
\tablewidth{0pt}
\tablecaption{Observations\label{table_obslog}}
\tablecolumns{15}
\tablehead{
\colhead{UT Date}
 & \colhead{Tel.\tablenotemark{a}}
 & \colhead{N\tablenotemark{b}}
 & \colhead{t\tablenotemark{c}}
 & \colhead{Filter}
 & \colhead{$\nu$\tablenotemark{d}}
 & \colhead{$R$\tablenotemark{e}}
 & \colhead{$\Delta$\tablenotemark{f}}
 & \colhead{$\alpha$\tablenotemark{g}}
 & \colhead{PA$_{-\odot}$\tablenotemark{h}}
 & \colhead{PA$_{-v}$\tablenotemark{i}}
 & \colhead{$\alpha_{pl}$\tablenotemark{j}}
 & \colhead{$m_R(R,\Delta,\alpha)$\tablenotemark{k}}
 & \colhead{$m_R(1,1,0)$\tablenotemark{l}}
 & \colhead{$Af\rho$\tablenotemark{m}}
}
\startdata
2012 Sep 11    & \multicolumn{4}{l}{\it Perihelion........................} & 0.0 & 2.411 & 1.753 & 21.4 & 262.1 & 243.5 &   6.3 & ... & ... & ... \\
2012 Oct 06    & PS1    &  1 &    40 & $r_{P1}$       &  7.4 & 2.414 & 1.540 & 14.4 & 273.5 & 244.1 &   6.8 & 19.6$\pm$0.1   & 15.9$\pm$0.3 & 10.1$\pm$2.9 \\
2012 Oct 08    & PS1    &  1 &    40 & $r_{P1}$       &  8.0 & 2.415 & 1.527 & 13.7 & 274.9 & 244.1 &   6.8 & 19.9$\pm$0.1   & 16.3$\pm$0.3 &  8.0$\pm$2.3 \\
2012 Oct 12    & Clay   & 38 &  2280 & $r'$           &  9.1 & 2.416 & 1.507 & 12.4 & 278.1 & 244.2 &   6.7 & 19.59$\pm$0.02 & 16.0$\pm$0.2 &  9.3$\pm$1.7 \\
2012 Oct 14    & UH2.2  &  1 &   300 & $R$            &  9.8 & 2.418 & 1.496 & 11.5 & 280.3 & 244.3 &   6.6 & 19.49$\pm$0.05 & 16.0$\pm$0.2 & 11.1$\pm$2.1 \\
2012 Oct 15    & UH2.2  &  2 &   600 & $R$            & 10.0 & 2.418 & 1.491 & 11.2 & 281.4 & 244.3 &   6.6 & 19.36$\pm$0.05 & 15.9$\pm$0.2 & 11.5$\pm$2.4 \\
2012 Oct 15    & FTS    &  8 &   480 & $R$            & 10.0 & 2.418 & 1.491 & 11.1 & 281.6 & 244.3 &   6.6 & 19.40$\pm$0.06 & 15.9$\pm$0.2 & 13.1$\pm$2.2 \\
2012 Oct 18    & Keck   &  4 &  1440 & $B$            & 10.9 & 2.419 & 1.479 & 10.0 & 285.3 & 244.4 &   6.4 & 20.17$\pm$0.02 & ...          & ... \\
               &        &  4 &  1200 & $R$            & ...   & ...   & ...  & ...  & ...   & ...   &   ... & 19.03$\pm$0.03 & 15.6$\pm$0.2 & 12.5$\pm$2.3 \\
2012 Oct 19    & Keck   &  4 &  1440 & $B$            & 11.2 & 2.419 & 1.475 &  9.7 & 286.8 & 244.4 &   6.4 & 20.23$\pm$0.02 & ...          & ... \\
               &        &  4 &  1200 & $R$            & ...   & ...   & ...  & ...  & ...   &   ... &   ... & 19.11$\pm$0.02 & 15.7$\pm$0.2 & 12.4$\pm$2.3 \\
2012 Oct 22    & UH2.2  & 14 &  4200 & $R$            & 12.0 & 2.421 & 1.466 &  8.5 & 292.0 & 244.5 &   6.2 & 19.01$\pm$0.02 & 15.7$\pm$0.2 & 11.4$\pm$2.1 \\
2012 Oct 25    & Baade  &  3 &   180 & $B$            & 12.9 & 2.422 & 1.459 &  7.6 & 298.2 & 244.5 &   6.0 & 19.98$\pm$0.06 & ...          & ... \\
               &        &  2 &   120 & $V$            & ...  & ...   & ...   & ...  & ...   & ...   &   ... & 19.33$\pm$0.04 & ...          & ... \\
               &        &  5 &   300 & $R$            & ...  & ...   & ...   & ...  & ...   & ...   &   ... & 19.05$\pm$0.03 & 15.8$\pm$0.2 & 11.4$\pm$2.1 \\
2012 Nov 8     & UH2.2  &  2 &   600 & $R$            & 17.0 & 2.431 & 1.455 &  5.3 &   0.5 & 244.8 &   4.8 & 18.98$\pm$0.04 & 15.8$\pm$0.2 & 13.0$\pm$2.4 \\
2012 Nov 9     & UH2.2  &  2 &  1200 & $R$            & 17.2 & 2.432 & 1.457 &  5.4 &   5.5 & 244.8 &   4.7 & 18.69$\pm$0.03 & 15.5$\pm$0.2 & 12.2$\pm$2.3 \\
2012 Nov 13    & UH2.2  & 26 &  7800 & $R$            & 18.4 & 2.434 & 1.467 &  6.3 &  22.6 & 244.8 &   4.2 & 18.57$\pm$0.02 & 15.3$\pm$0.2 & 12.8$\pm$2.4 \\
2012 Nov 14    & UH2.2  & 14 &  4200 & $R$            & 18.7 & 2.435 & 1.470 &  6.5 &  26.1 & 244.8 &   4.1 & 18.76$\pm$0.02 & 15.5$\pm$0.2 & 12.7$\pm$2.4 \\
2012 Nov 22    & Lowell &  4 &  2400 & $R$            & 21.0 & 2.441 & 1.502 &  9.1 &  44.9 & 244.7 &   3.1 & 19.11$\pm$0.03 & 15.7$\pm$0.2 & 13.1$\pm$2.4 \\
2012 Nov 23    & Lowell &  2 &  1400 & $R$            & 21.3 & 2.442 & 1.507 &  9.5 &  43.9 & 244.7 &   3.0 & 18.95$\pm$0.04 & 15.5$\pm$0.2 & 15.5$\pm$2.9 \\
2012 Dec 18    & UH2.2  & 30 &  9000 & $R$            & 28.3 & 2.467 & 1.712 & 17.7 &  64.5 & 244.1 & --0.1 & 19.47$\pm$0.02 & 15.4$\pm$0.2 & 10.6$\pm$2.0 \\
2012 Dec 19    & UH2.2  & 49 & 15000 & $R$            & 28.6 & 2.468 & 1.723 & 17.9 &  64.8 & 244.1 & --0.2 & 19.48$\pm$0.02 & 15.4$\pm$0.2 & 11.0$\pm$2.0 \\
2012 Dec 20    & Lowell &  4 &  4800 & $R$            & 28.9 & 2.470 & 1.734 & 18.2 &  65.1 & 244.1 & --0.3 & 19.82$\pm$0.04 & 15.7$\pm$0.2 &  9.5$\pm$1.8 \\
2013 Jan 08    & UH2.2  & 10 &  8400 & $R$            & 34.2 & 2.493 & 1.962 & 21.6 &  69.5 & 243.7 & --2.1 & 20.40$\pm$0.02 & 15.9$\pm$0.2 &  8.2$\pm$1.5 \\
2013 Feb 04    & SOAR   &  1 &   600 & $R$            & 41.4 & 2.531 & 2.334 & 22.9 &  73.3 & 243.9 & --3.5 & 21.41$\pm$0.05 & 16.5$\pm$0.2 &  6.0$\pm$1.1 \\
2015 Jun 27    & \multicolumn{4}{l}{\it Aphelion..........................} & 180.0 & 3.896 & 3.656 & 15.0 & 114.2 & 296.1 &   0.5 & ... & ... & ... \\
2018 Apr 12    & \multicolumn{4}{l}{\it Perihelion........................} &   0.0 & 2.402 & 3.392 &  3.1 & 134.3 & 243.4 & --2.9 & ... & ... & ...
\enddata
\tablenotetext{a}{Telescope.}
\tablenotetext{b}{Number of exposures.}
\tablenotetext{c}{Total integration time, in s.}
\tablenotetext{d}{True anomaly, in degrees.}
\tablenotetext{e}{Heliocentric distance of object, in AU.}
\tablenotetext{f}{Geocentric distance of object, in AU.}
\tablenotetext{g}{Solar phase angle (Sun-object-Earth), in degrees.}
\tablenotetext{h}{Position angle of the antisolar vector, in degrees East of North.}
\tablenotetext{i}{Position angle of the negative velocity vector, in degrees East of North.}
\tablenotetext{j}{Orbit plane angle, in degrees.}
\tablenotetext{k}{Mean apparent $R$-band magnitude.}
\tablenotetext{l}{Absolute $R$-band magnitude (at $R=\Delta=1$~AU and $\alpha=0^{\circ}$), assuming solar colors and IAU $H,G$ phase-darkening where $G=0.15$. Listed uncertainties are dominated by the estimated uncertainty in $G$.}
\tablenotetext{m}{Dust contribution (computed using $5\farcs0$ photometry apertures), as parameterized by \citet{ahe84}, in cm.}
\end{deluxetable}

\end{document}